\documentclass[rnote]{aa}       
\usepackage{graphicx}
\usepackage{amssymb}
\usepackage{longtable}
\usepackage{supertabular}
\usepackage{natbib} 

\begin{document}

\title{Fermi~I particle acceleration in converging flows mediated by magnetic reconnection}

\author{V. Bosch-Ramon \inst{1}}

\authorrunning{Bosch-Ramon, V.}

\titlerunning{Fermi~I particle acceleration in converging flows}

\institute{Dublin Institute for Advanced Studies, 31 Fitzwilliam Place, Dublin 2, Ireland; valenti@cp.dias.ie
}

\offprints{V. Bosch-Ramon, \email{valenti@cp.dias.es}}

\date{Received <date> / Accepted <date>}

\abstract
{Converging flows with strong magnetic fields of different polarity can accelerate particles through magnetic
reconnection. If the particle mean free path is longer
than the reconnection layer is thick, but much shorter than
the entire reconnection structure, the particle will mostly interact with the incoming flows potentially with a 
very low escape probability.}   
{We explore, in general and also in some specific scenarios, 
the possibility of particles to be accelerated in a magnetic reconnection layer by interacting only 
with the incoming flows.}  
{We characterize converging flows that undergo magnetic reconnection, 
and derive analytical estimates for the 
particle energy distribution, acceleration rate, 
and maximum energies achievable in these flows. We also discuss a scenario, based 
on jets dominated by magnetic fields of changing polarity, in which this mechanism may operate.}  
{The proposed acceleration mechanism operates if the reconnection layer is much thinner than its transversal
characteristic size, and the magnetic field has a disordered component. Synchrotron losses may prevent electrons from
entering in this acceleration regime. The acceleration rate should be faster, 
and the energy distribution of particles harder 
than in standard diffusive shock acceleration. The interaction of obstacles with the innermost region of jets 
in active galactic nuclei and microquasars may be suitable sites for particle acceleration in converging flows.}
{} 

\maketitle

\section{Introduction} \label{intro}

The interaction of outflows with themselves or their environment usually leads to particle acceleration
\citep[e.g.][]{rie07}, and the most common acceleration mechanism in these circumstances is thought to be diffusive shock
(Fermi~I) acceleration  \citep[e.g.][]{kri77,bell78a,bell78b,bla78,dru83}. This process works well when a plasma with a weak
disordered magnetic field ($B$) suffers a strong shock. The Fermi~I acceleration efficiency depends on the energy gain when
particles cross the shock, and on the escape probability downstream the shock, which is related to the shock
compression ratio $R=v/v'$, $v'(v)$ being the post(pre)-shock velocity.

In strong non-relativistic adiabatic shocks in the test-particle approximation $R$ equals 4, 
and the resulting energy distribution
is $Q(E)\propto 1/E^2$ \citep[e.g.][]{dru83}. In radiative shocks, $R\gg 4$, but densities become very high and energy losses
and damping of magnetic irregularities may limit the efficiency of the acceleration process (e.g. \citealt{ddk96},
\citealt{rkd07}; see nevertheless Sect.~\ref{accrad}). For plasmas that carry 
a strong perpendicular $B$-component, $R$ is
$\sim 1$, and Fermi~I acceleration is suppressed. However, if the $B$-field has reversals, energy can be dissipated via
magnetic reconnection in regions where the charge density is not sufficiently high 
to sustain different polarity $B$-lines
\citep[e.g.][]{par57,pet64,spe65,son71,zen01}, allowing for $R\gg 1$ \citep[e.g.][]{dru12}.

\begin{figure}
\includegraphics[width=90mm,angle=-0]{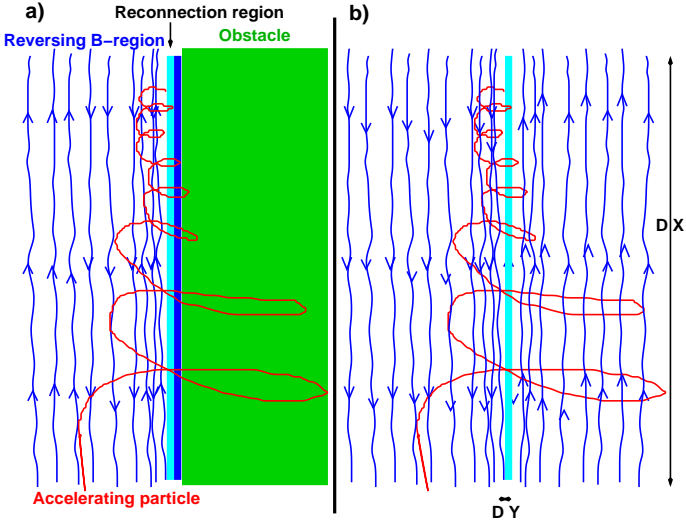}
\caption{Sketch of the considered scenario: a) the interaction of one 
flow carrying different polarity $B$-lines 
with an obstacle (see Sect.~\ref{disc}); b) the interaction of two flows with different polarity $B$-lines.}
\label{sk}
\end{figure}

Studies show that magnetic reconnection can accelerate the bulk of particles up to the pre-reconnection 
Alfv\'enic speed, with
some particles reaching even higher energies \citep[e.g.][]{zen01,lyu08,kdl11}. Magnetic reconnection may also lead to
the formation of strong shocks, accelerating particles through the standard Fermi~I mechanism in an otherwise weakly
compressive flow \citep[e.g.][]{for88,bla94,sir11a}. Fermi~I particle acceleration may operate in the reconnection
region itself \citep[e.g.][]{gou05,gia10,dru12}, since particles can bounce back and forth between the layer and the
converging flows. The mean free path $\lambda$ of these particles can eventually overcome the layer thickness, $\Delta Y$,
and then particles will not interact significantly with the reconnection layer, 
but with the perpendicular $B$-lines
of the converging flows \citep[][]{gia10}. In this case, the
advection escape probability becomes formally zero because the flows do not move away from the 
reconnection layer. This is true as long as the mean free path of the particle
parallel to $B$, $\lambda_\parallel$, is much shorter than the characteristic size of the entire reconnection region, $\Delta
X$ (to order the magnetic field reversal scale). Escape in the opposite direction to the
incoming flow can be neglected unless the region
extension in this direction is $\ll \Delta X$. A schematic picture of this scenario is presented in Fig.~\ref{sk}, 
which shows two
cases: the interaction of a flow carrying different polarity $B$-lines with an obstacle (described
in Sect.~\ref{disc}); and the interaction of two flows with different polarity $B$-lines. As shown in the figure, particles
can spiral back and forth between the two sides of the reconnection layer until they diffuse away from the region.

In this note, we analytically explore the Fermi~I mechanism when particles interact with converging flows in the
null-escape probability regime. We assume that particles can effectively diffuse through the incoming flows. We focus on
the case when magnetic reconnection is the mechanism that leads to $R\gg 1$. 
We derive in Sect.~\ref{acc} analytical
estimates for the acceleration rate, the 
maximum energy and energy distribution of the accelerated particles, and present in
Sect.~\ref{disc} an illustrative case in the context of galactic and extragalactic jets. For simplicity, the discussions are
restricted to Newtonian flows, although our conclusions qualitatively apply to the relativistic case as well. 
For some magnitudes, we will adopt the convention $Q_x=(Q/10^x\,{\rm cgs})$.

\section{Particle acceleration in converging flows}\label{acc}

In converging flows, particles slowly diffuse perpendicularly to the $B$-lines, with $\lambda_{\perp}\sim r_g/\chi$,
where $\chi$, with value $\ge 1$, is the total to the disordered magnetic energy density ratio. 
Remarkably, 
a disordered $B$-component can enhance the reconnection rate and the reconnecting flow velocity $v$, as shown for instance 
in \cite{laz99}.
The following
limit for Fermi~I to operate between the converging
flows can be imposed: 
in the current sheet, $\lambda_\perp>\Delta Y$, i.e. $E_{\rm min}\sim \chi\,q\,B_{\rm r}\,\Delta Y$, 
where $B_{\rm r}$ is
the $B$-field in there. Although one can expect that $B_{\rm r}\ll B$ close to the center of the current sheet, 
we conservatively assume that over the whole reconnection layer $B_{\rm r}\sim B$. On
the  other hand, the $B$-lines become more and more entangled close to the reconnection layer, so one can set there 
$\chi=1$. The previous condition becomes
\begin{equation}
E>E_{\rm min}\sim 0.3\,B_{\rm 0}\,\Delta Y_9\,{\rm TeV}.
\end{equation}
In addition, to be efficiently accelerated, 
particles in the incoming flows should not drift too early along the $B$-lines, parallel to the layer, 
out of the reconnection region 
(cross-field diffusive escape will be slower by $1/\chi^2$). 
This is a sort of Hillas limit: $\lambda_\parallel\sim \chi\,r_{\rm g}<\Delta X$, or
\begin{equation}
E_{\rm max}^{\rm H}\sim q\,B\Delta X/\chi\approx 3\,B_{\rm 0}\Delta X_{10}\chi^{-1}{\rm TeV}.
\label{maxh}
\end{equation}
This gives the maximum dynamical range 
\begin{equation}
E_{\rm max}^{\rm H}/E_{\rm min}\sim 10\,\chi^{-1}\,(\Delta X_{10}/\Delta Y_9)\,.
\label{dinh}
\end{equation}
\cite{gia10} accounted for the electric field between the sides of the reconnection layer: $\epsilon\sim v\,B$. The
$\epsilon$-field accelerates the particles perpendicularly to $B$ and $v$, with $E_{\rm max}\sim (v/c)qB\Delta X$. 
The fraction of time spent by particles within the reconnection layer during an acceleration cycle is 
$\Delta Y/c\,t_{\rm cycle}$, and thus this effect will be important at $E\sim E_{\rm min}$ or
for very ordered fields.

From Eqs.~(\ref{maxh}--\ref{dinh}) one can see that the quantities $\chi$ and $\Delta X/\Delta Y$ are crucial for the
efficiency of the discussed acceleration mechanism. The derivation of $\chi$ requires a detailed treatment of the plasma, but
as mentioned, given the dramatic change of the $B$-geometry, close to the reconnection layer $\chi\sim 1$. On scales of 
$\Delta X$, and given the high plasma characteristic speeds, the disordered $B$-field generated in the reconnection layer may
propagate to the converging flows, although such a $B$-component may also have an external origin. To estimate $\Delta
X/\Delta Y$, we briefly discuss the flow dynamics below.

\subsection{Flow dynamics} 

\begin{figure}
\includegraphics[width=80mm,angle=-0]{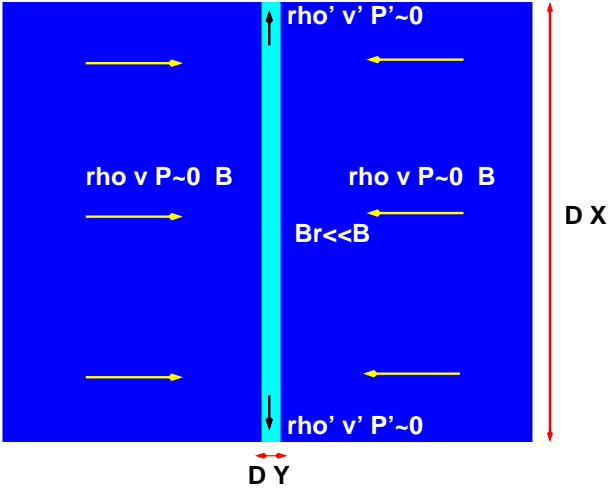}
\caption{Sketch of the converging and the outgoing flows in the reconnection region.}
\label{sk2}
\end{figure}

The plasma that enters, flows along, and eventually leaves
the reconnection layer, as shown in Fig.~\ref{sk2}, can be described by the particle and 
energy flux conservation equations: 
\begin{equation}
A\rho\,v=A'\rho'\,v'\,\,\,\,{\rm and}
\label{cons1}
\end{equation}
\begin{equation}
A\,v\,B^2/8\pi=A'\rho'\,v'^{3}/2\,,
\label{cons2}
\end{equation}
where unprimed/primed quantities correspond to the incoming/outgoing flows, $\rho$ is the 
density, $v$ the
velocity, and $A\sim \Delta X^2$ and $A'\sim 4\Delta X\Delta Y$ are the in- and out-flow areas, respectively.
The flow pressure is assumed to be
negligible in the converging flows because of strong $B$-dominance. When leaving the region,
under negligible external pressure 
and reasonable geometries, the flow kinetic energy becomes dominant and 
pressure can be neglected as well. 

Given the symmetry of the problem, there is no momentum conservation equation, and the conditions deep within
the reconnection layer are hard to determine. Equations \ref{cons1} and \ref{cons2} yield $v'\approx v_{\rm
A}=\sqrt{B^2/4\pi\rho}$ and thus $\rho'\approx \rho\,(A/A')(v/v_{\rm A})$, but one has to make additional 
assumptions to obtain $\Delta X/\Delta Y$. An estimate may be derived assuming one-dimensional adiabatic
compression of $B_{\rm r}$, the remaining field after reconnection in the layer, until it balances the
incoming field pressure. This gives $\rho'/\rho\sim (B/B_{\rm r})$ and therefore an area ratio $A'/A=(B_{\rm
r}/B)(v/v_{\rm A})$, or $\Delta X/\Delta Y\sim 4(B/B_{\rm r})(v_{\rm A}/v)$. An upper limit for $\Delta
X/\Delta Y$ would come from $\Delta Y\sim <r_{\rm g}>$, where $<r_{\rm g}>$ is the gyroradius of the average
particle in the reconnection layer. A lower limit may be derived assuming equilibrium between the upstream
magnetic pressure and the thermal pressure inside the layer of thickness $\Delta Y$, implying $(\Delta
X/\Delta Y)\sim 4(v_{\rm A}/v)$. One can therefore conclude that the dynamical range may span several orders of
magnitude in energy, but it needs an efficient magnetic-to-kinetic energy transfer \citep[which may be
reasonable,  as suggested in][]{dru12}.

\subsection{Acceleration and radiation processes}\label{accrad}

Like in standard diffusive shock acceleration, 
particles gain energy by moving back and forth  between the incoming flows as $\Delta E/E\sim v/c$ per cycle. 
In addition,
since there is no advection directed outwards, the probability to cross back to the other side of the
reconnection layer becomes 1 after covering few $\lambda_\perp$ in each incoming flow, i.e.  
$t_{\rm cycle}\sim (k/\chi)\,r_{\rm g}/c$, with $k\sim
10$. This $t_{\rm cycle}$ gives an acceleration rate
\begin{equation}
\dot{E}_{\rm acc}\sim \Delta E/t_{\rm cycle}=(\chi/k)\,(v/c)\,q\,B\,c\,,
\label{acce}
\end{equation}
and the zero advection escape probability renders a particle energy distribution
$Q(E)\propto 1/E$ 
up to energies a few times lower than $E_{\rm max}^{\rm H}$, at which the distribution drops very
quickly. Note that this mechanism is faster in $\dot{E}_{\rm acc}$ 
($(v/c)$ vs $(v/c)^2$) and yields a harder
$Q(E)$ ($1/E$ vs $1/E^2$) than
standard Fermi~I.
Such a hard distribution of accelerated particles can become dominant in pressure.
This should smooth the $v$ profile 
and affect $Q(E)$ (as in non-linear Fermi~I acceleration; e.g. \citealt{dru83}).

Radiation losses can stop the acceleration at $E<E_{\rm max}^{\rm H}$. Typically, for protons these losses are 
fairly
inefficient, whereas for electrons synchrotron cooling \citep[e.g.][]{blu70} tends to be important, 
which renders a maximum energy
\begin{equation}
E_{\rm max}^{\rm sy}\approx 60\,(\chi/k)^{1/2}(v/c)^{1/2}B_{\rm 0}^{-1/2}\,{\rm TeV}.
\label{maxsy}
\end{equation}
In environments with very dense matter or photon fields, other cooling processes might be relevant, 
such as relativistic
Bremsstrahlung and inverse Compton (IC) for electrons \citep[e.g.][]{blu70}, and proton-proton collisions ($pp$), 
synchrotron and photomeson
production for protons \citep[e.g.][]{kel06,aha00,kel08}. 
Diffusive escape parallel to $B$ can also be more restrictive than the Hillas limit: 
\begin{equation}
E_{\rm max}^{\rm di}\approx (3v/2c\,k)^{1/2}q\,B\Delta X\approx 4\,(v/c\,k)^{1/2}B_{\rm 0}\Delta X_{10}\,{\rm TeV}. 
\end{equation}
The corresponding dynamical ranges are
\begin{equation}
E_{\rm max}^{\rm sy}/E_{\rm min}\sim 200\,(\chi/k)^{1/2}(v/c)^{1/2}B_{\rm 0}^{-3/2}\Delta Y_9^{-1}\,\,\,\,{\rm and}
\end{equation}
\begin{equation}
E_{\rm max}^{\rm di}/E_{\rm min}\sim 10\,(v/c\,k)^{1/2}\,\Delta X_{10}\,\Delta Y_9^{-1}.
\end{equation}

In general, to accelerate particles to TeV energies, at least mildly relativistic $v_{\rm A}$-values will be required. In
addition, for electrons, if converging flow acceleration is to be efficient, 
$\Delta Y_9\lesssim B_{\rm 1.5}^{-3/2}$. This will
typically require in astrophysical sources either very high 
compression ratios, or very low magnetic fields, the latter implying a very large structure 
(e.g. magnetized turbulent ISM) if the process is to have observable radiative effects. For protons, the Hillas and diffusive
limits lead to the more reasonable conditions $\Delta X>\chi\Delta Y$ and $>(v/c\,k)^{-1/2}\Delta Y$, respectively. Note nevertheless that
electrons remaining within the reconnection layer may still be 
accelerated by electric fields generated in the reconnection
process, resulting in energies $\lesssim E_{\rm min}$. In this case, the accelerated energy distribution
could also be
$Q(E)\propto 1/E$ \citep[e.g.][]{zen01}. 

Synchrotron emission from electrons injected with $Q(E)\propto 1/E$ and cooled through this process would have a spectral
energy distribution $\nu\,L_\nu\propto \nu^{1/2}$. On the other hand, proton radiation would show $\nu\,L_\nu\propto \nu$ for
synchrotron (or $\propto\nu^{1/2}$ in saturation) and for $\pi^0$-decay from $pp$ and photomeson
production (above threshold). The last two processes also generate secondary $e^\pm$ pairs from $\pi^\pm$-decay with spectra and
luminosities similar to those of the photons from $\pi^0$-decay. These pairs  will predominantly cool through
synchrotron, like primary electrons, also with $\nu\,L_\nu\propto \nu^{1/2}$ but peaking at much higher energies.

Before exploring the case of magnetized jets, we note that our results may also 
be applied to weakly magnetized flows
with radiative shocks. Particle acceleration between the upstream and the cooled downstream would require 
sufficiently energetic 
particles to cross the adiabatic region of the post-shock flow and reach the cooled downstream medium. Another condition
would be that the particle cooling/escape timescales were longer everywhere than the acceleration
one. In this context, young stellar object jet shocks, or the dense winds of two massive stars colliding, may be
sources worthy of being investigated. 

\section{The case of magnetized jets}\label{disc}

Powerful astrophysical outflows, such as 
pulsar winds, and microquasar and AGN jets, are expected to be magnetically dominated
at their formation and acceleration zones \citep[e.g.][]{cor90,bog99,bes06,kom07}. 
In pulsar winds, magnetic reconnection may
take place naturally in a current sheet 
that originated at the wind equatorial region \citep{cor90,lyu01}.  In relativistic jets,
a $B$-field with reversals could come from the accretion disc \citep[e.g.][]{bb11}, but magnetic dissipation may need to be
triggered through jet acceleration, which could drive magnetic reconnection through the Kruskal-Schwarzschild instability
\citep[see][]{lyu10}. A similar though more extreme effect can be expected if an obstacle is entrained by the jet
\citep[e.g.][]{hub06,abr09,abr10,bab10,pb12,bpb12,babkk10} with a dominant $B$-field with polarity reversals of size $\Delta X$.
Under the ongoing magnetic reconnection, the incoming jet material and the obstacle itself can play the role of converging
flows with a high $\Delta X/\Delta Y$ ratio. To avoid the suppression of the acceleration process, 
the shocked obstacle must
fulfil the following conditions: high inertia, to avoid quick dragging by the jet; not too high density, 
to avoid fast
radiation cooling; and a mean free path much shorter than the obstacle size ($D_{\rm o}$), 
to avoid fast particle escape. 

For a jet/obstacle interaction at $z_{\rm j}\sim 100\,R_{\rm Sch}$ from the central object, where $R_{\rm Sch}\approx 3\times
10^{13}\,(M/10^8\,M_\odot)\,{\rm cm}$, a jet width $D_{\rm j}\sim 0.1\,z_{\rm j}\sim 3\times 10^{14}\,(M/10^8\,M_\odot)\,{\rm
cm}$ and magnetic field $B\sim 10^3\,L_{\rm j,44}^{1/2}\,(M/10^8\,M_\odot)^{-1}\,{\rm G}$, and 
$D_{\rm o}\gtrsim \Delta X$, the luminosity budget available for dissipation will be
$L_{\rm d}\sim (D_{\rm o}/D_{\rm j})^2\,L_{\rm j}\le L_{\rm j}$, 
where  $L_{\rm j}$ is the jet power. Assuming $\Delta X\sim R_{\rm Sch}$, the electron and proton maximum energies 
will be 
$E_{\rm max}^e\sim 1\,(M/10^8\,M_\odot)^{1/2}\,L_{\rm j,44}^{-1/4}\,{\rm TeV}$,
and 
$E_{\rm max}^p\sim 10^7\,L_{\rm j,44}^{1/2}\,{\rm TeV}$. 

Given the strong magnetic fields, the condition $\Delta Y_9\lesssim B_{\rm 1.5}^{-3/2}$ may not hold; if so, electrons could
only be accelerated within the reconnection layer. Thus, in AGN jet/obstacle interactions, magnetic reconnection
itself  could potentially accelerate electrons up to TeV energies. Protons, on the other hand, 
could be accelerated in
the converging flows\footnote{Even if jets were formed only by $e^\pm$ pairs, protons may be entrained from the
environment, e.g. the obstacle itself.} to ultra high energies or even beyond depending on $\Delta X$, $L_{\rm
j}$, and $D_{\rm o}$ \citep[see][for similar results]{gia10}. Electrons would yield hard synchrotron radiation
peaking around 10~MeV. Proton synchrotron emission, of hard spectrum and peaking around 100~GeV, may be efficient
as well under these conditions \citep[see, e.g.,][]{babkk10}. 
Emission through $pp$ inside the obstacle could be
also significant \citep{bab10}, as well as photomeson production in very bright AGN. Secondary very energetic
$e^\pm$ pairs from $\pi^\pm$-decay would radiate ultra high-energy photons through synchrotron emission that
could pair-create in the jet radio fields on pc-scales, being reprocessed as synchrotron and IC photons of
lower energies.

In microquasar jets, where, say, $M\sim 10\,M_\odot$ and $L_{\rm j}\sim 10^{36}$~erg~s$^{-1}$, magnetic reconnection in the
jet/obstacle boundary at  $z_{\rm j}\sim 100\,R_{\rm Sch}$ could lead to synchrotron MeV flares produced by GeV electrons. A
minor synchrotron self-Compton component in GeV may be also expected. 
PeV protons may be accelerated in the converging flows,
and if they met a nearby thick target, for instance 
the obstacle itself or a dense field of energetic photons, they could produce hard
emission peaking at $\sim 100$~TeV through $pp$ or photomeson production, plus hard TeV synchrotron radiation from the
subsequent $e^\pm$ pairs. In compact and powerful objects photon-photon absorption would be severe, radiation being 
re-emitted
as soft gamma~rays via synchrotron emission. It is not clear which kind of obstacle may be present in the innermost regions
of a microquasar jet, although in the (unlikely) event that the jet remained strongly magnetized up to the binary scales
($\gg 100\,R_{\rm Sch}$), a clumpy stellar wind could provide those obstacles \citep{abr09}. 

In both microquasars and AGN, the reconnection process can be efficient as long as the obstacle has a velocity different from
that of the flow. In this process, for $D_{\rm o}\sim D_{\rm j}$, a significant fraction of $L_{\rm j}$ may be released, with
potential fluxes at high energies $F\sim 10^{-10}\,{\rm erg~cm}^{-2}\,{\rm s}^{-1}$ $\times L_{\rm d,44}\,(d/100\,{\rm
Mpc})^{-2}$ for AGN, and $\times L_{\rm d,36}\,(d/10\,{\rm kpc})^{-2}$ for microquasars. This radiation would be roughly
isotropic, but for obstacles small and light enough to be accelerated by the jet, the emission would become 
progressively
Doppler-boosted and enhanced for specific viewing angles. The obstacle may expand, enhancing the radiation, and possibly
fragment while accelerated. All this should lead to complex spectral and variability patterns \citep[e.g.][]{bab10,babkk10}.
The typical timescale of the whole interaction process will be the time the obstacle remains as such \citep[see,
e.g.,][]{bpb12}, whereas the reconnection events will have a variability timescale $\sim \min[D_{\rm
o},\Delta X]/v$, where $v\lesssim c$. 

\begin{acknowledgements}
We thank an anonymous referee for constructive and useful comments and suggestions.
We are grateful to Maxim Barkov, Luke Drury and Dmitry Khangulyan for fruitful discussions.
The research leading to these results has received funding from the European
Union Seventh Framework Program (FP7/2007-2013) under grant agreement
PIEF-GA-2009-252463. V.B.-R. acknowledges support by the Spanish 
Ministerio de Ciencia e Innovaci\'on
(MICINN) under grants AYA2010-21782-C03-01 and FPA2010-22056-C06-02.  
\end{acknowledgements}

\bibliographystyle{aa}
\bibliography{text}
\end{document}